# Frequency-based Ultrasonic Backscatter Modulation for Passive Sensing Applications

Muhammad Junaid Akhtar, Alp Timuçin Toymuş, Kıvanç Esat, and Levent Beker

***Abstract*—Ultrasonic backscatter communication has gained popularity in recent years where piezoceramic resonators are used as acoustic antennas. Currently, backscatter communication revolves around amplitude modulation of the echo signal by the sensor, which can be modelled as a variable shunt impedance. Amplitude based sensing is prone to high levels of noise, motion artifacts, and has low efficiency with respect to power usage and bandwidth. To overcome these shortcomings, we present here, a frequency-based sensing method for ultrasonic backscatter communication. This system exhibits a frequency shift in the ultrasonic sensor's parallel resonance when its shunt capacitance is varied, similar to an inductive near-field link. The concept is simulated using an equivalent end-to-end model in SPICE that matches well with the experimental observation. To monitor the resonance frequency shift in the experiments a variety of methods are employed including ringdown measurements, chirp spectroscopy, bode analysis and phase locked loop. All the methods provide substantial proof for feasibility of frequency-based sensing in ultrasonic backscatter communication enabling passive sensor motes for sensing applications.**

## I. INTRODUCTION

ULTRASOUND is being widely used for imaging, obstacle detection, and power transfer in various airborne and medical applications. Another potential venue that could utilize ultrasound is wireless sensing nodes (USNs). Conventional inductor-capacitor (LC) based passive sensor nodes have been used to develop battery-free passive sensing for variety of applications ranging from wireless structural/environmental monitoring to implanted medical devices [1]. However, short communication range and misalignment effects limit potential applications that demand mobility, long range, and robustness. Recently, utilization of ultrasonic backscattered data transmission for remote sensing of physiological signals such as pressure and temperature gained attention [1]. Ultrasonically powered sensor nodes have been shown to be a viable solution for various battery-free applications including deep tissue therapy and biosensing [2]. Extraction of data and communication with the sensor is a key design aspect of ultrasonic sensor nodes. Communication with USN has been the aim of many research efforts [3-4] .

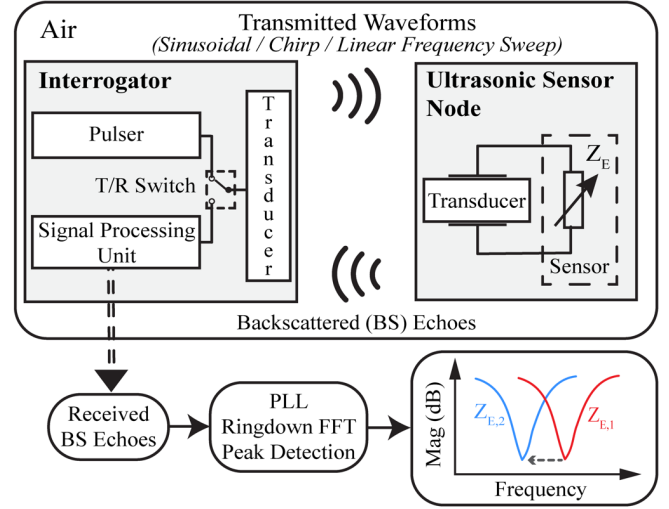

Fig. 1. Backscatter ultrasonic (US) airborne transducer communication

Basically, a USN consists of an ultrasonic communication device such as a piezo resonator attached to a sensor. In simplest form, the sensor can be modelled as a variable shunt impedance $Z_E$, as most sensors represent any measured parameter as a change in its impedance value. This impedance can purely be resistive, as in case of temperature sensor, or capacitive, such as a pressure sensor [5]. In this configuration, the piezo functions as an acoustic antenna, which enables the sensor to receive ultrasonic waves from an external transducer. A change in the electrical load $Z_E$ alters the acoustic reflection coefficient of the piezo, which leads to amplitude modulation of the backscattered signal (echo). The piezo then transmits this modulated signal back to the external transducer, where the signal is processed to extract the data (i.e., sensor response). In existing methods, the shunt impedance $Z_E$ is used to modulate the amplitude of backscattered echo [5]. Considerable research has been done to determine the sensitivity of the acoustical reflection coefficient to variations in electrical load $Z_E$ [5], [6]. However, amplitude modulation has drawbacks such as low power usage and bandwidth efficiency, and it is more prone to noise and misalignment effects. On the other hand, frequency modulation does not have these shortcomings and has proven to be more robust and is a widely accepted communication method. In this paper, we propose a frequency modulated backscatter communication for USNs. First, using LTSPICE, we simulate the end-to-end equivalent circuit model of the communication channel by modelling the shunt impedance $Z_E$ as a variable capacitor. We observe the frequency modulation of the backscattered echo by the variable capacitance. After determining feasibility from simulations, we perform the experiments using ultrasonic transducers in the air domain. Different methods are employed to quantify the frequency modulation of backscattered echo experimentally, including ringdown signal analysis, Chirp, linear frequency sweep, and phase locked loop (PLL).

Fig. 1 summarizes the proposed approach[7]. The interrogator generates the ultrasonic waves with a pulser that excites an ultrasonic transducer through a transmit/receive (T/R) switch. These ultrasonic waves are transmitted to the USN in air, where they are received by an ultrasonic transducer. The variable capacitor that behaves as the sensor, varies the electrical resonance of the USN and modulates the signal it receives from the US transducer. This modulated signal, referred to as backscattered echo, is transmitted back to the interrogator. The backscattered echo signal is sent to a signal processing unit through the T/R switch. Different signal processing techniques are applied on the backscattered echo signal to estimate parallel resonance frequency shift including ringdown signal analysis, chirp spectroscopy, Bode analysis and frequency tracking using a phase locked loop.

The manuscript is organized as follows: Description of channel equivalent circuit model is given in Section II. SPICE (Simulation Program with Integrated Circuit Emphasis) simulation of communication channel and its results are discussed in Section III. In Section IV, experimental setup of the concept is presented. The results of different excitations and signal processing techniques are discussed in Section V, and Section VI concludes the paper.

## II. CHANNEL EQUIVALENT CIRCUIT MODEL

To model a piezoelectric transducer, different equivalent circuit models have been proposed over the years. Mason [8] proposed an exact equivalent circuit by using an electrical port and two acoustic ports. But the negative capacitance in this model is non-physical and difficult to model in simulation tools [9]. Redwood [10] and KLM [11] models removed this negative capacitance, but they still use a frequency dependent transformer turn ratio that cannot be easily simulated. Leach model avoids both issues [12] and employs controlled sources instead of a frequency-dependent transformer. These characteristics make Leach model easier to implement in SPICE, therefore; it was utilized for simulations as shown in Fig. 2.

Here, $T_1$ is a lossless transmission line with characteristic impedance of the transducer ($Z_C$) equal to the radiation acoustic impedance of the transducer i.e., $Z_C = Z_0A$ where $Z_0$ is the specific acoustic impedance of the transducer material, and A is the cross-sectional area of the US transducer. The node labels E, B, and F, respectively, denote the electrical, acoustical back, and front ports. $C_0$ is the capacitance of the transducer. The independent voltage sources $V_1$ and $V_2$, are zero valued sources which are used as ammeters in the circuit. The depended source $F_1$ has value $hC_0$ and $F_2$ has value h, where h is the piezoelectric constant. The voltage across dependent source $E_1=V(4)$, where V(4) is the voltage at labelled node 4. Values of $R_1$ and $C_1$ are fixed at $R_1=1 k\Omega$ and $C_1=1$ F.

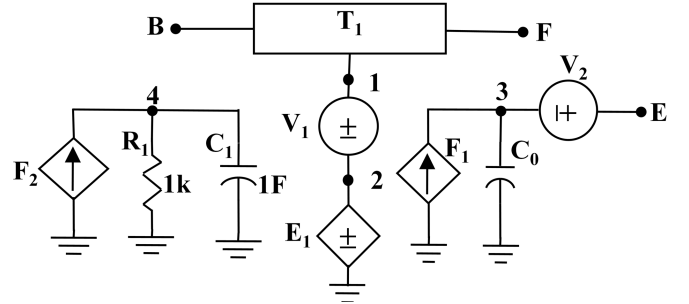

Fig. 2. Leach model for thickness mode transducer [11].

To simulate the backscatter response of the channel, we use an equivalent circuit model as shown in Fig. 3(a). Identical US transducers are used in the USN and interrogator. Both of the US transducers are used in the thickness vibration mode. The second acoustical ports of both transducers in Fig. 3(a) are terminated by the radiation acoustic impedance of the backing layer. Backing layer impedances of USN ($Z_B$) and interrogator ($Z_{Bext}$), are both set to zero because air is used as backing layer for both transducers.

Both US transducers are placed in the same medium i.e., air, which eliminates the requirement for any matching layer. Air gap between both transducers is 2 mm. To account for path loss, the propagation medium (air) is assumed to be a lossy transmission line. The parameters of the lossy transmission line are found as follows [13]:

$$R_a = 2\rho_a v_a A\alpha \tag{1}$$

$$L_a = A\rho_a \tag{2}$$

$$C_a = \frac{1}{A\rho_a v_a^2} \tag{3}$$

Where $\rho_a$ is the density of air and $v_a$ is the speed of sound in air. A is the cross-sectional area of the US transducer and $\alpha$ is the co-efficient of attenuation in air. All the parameters of the Leach circuit model can be calculated once the transducer material, geometry, and type of resonance mode are known. Parameters used in the simulation are listed in Table I.

## III. SIMULATION

For any piezoelectric transducer, there are two resonance frequencies. The frequency where the transducer has minimum impedance is called series (resonance) frequency, and the frequency where impedance is maximum is called parallel (anti-resonance) frequency. The addition of a shunt capacitive load decreases both the series and parallel resonance frequencies of a piezoelectric transducer [14]. Both the resonance frequencies keep on decreasing with increase in load capacitance. When the load capacitance become large, it's effect on the serial and parallel resonance frequencies becomes negligible. By looking

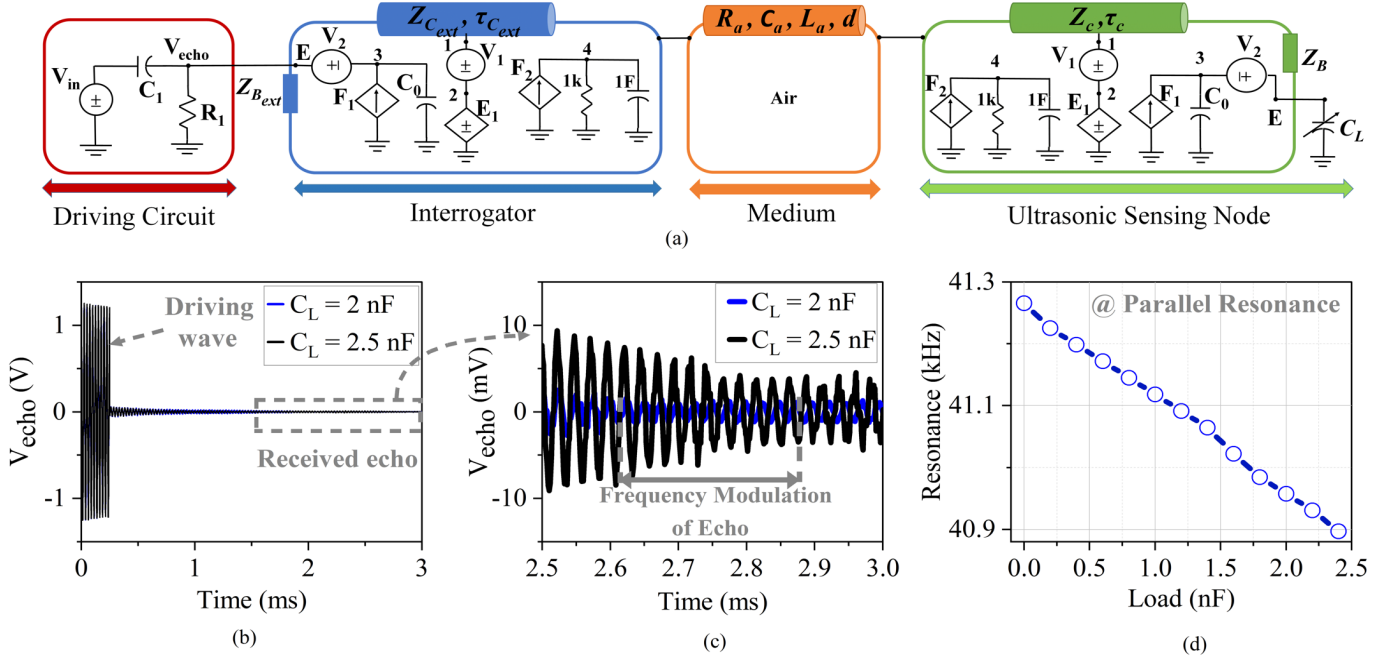

Fig. 3 a) Backscatter communication equivalent circuit model used in simulations. Simulated (b) transient response of channel (c) echo voltage frequency modulation (d) parallel resonance frequency vs. capacitive load, $C_L$.

at impedance plots of the transducer with range of shunt load capacitance used in this work, it was observed that increase of load capacitance changed series resonance by a very small amount as compared to a noticeable change in parallel resonance. The resonance frequency shift is more sensitive to change in load capacitance around the parallel resonance. Therefore, to observe frequency shift with varying electrical load capacitance, we excite the signal at the parallel resonance frequency of the transducer. The simulation of the model shown in Fig. 3(a) was performed in LTSPICE. The interrogator US transducer is excited by 10 cycles of sinewave at the parallel resonant frequency of the ultrasonic transducer, as shown in Fig. 3(b). After some time, the backscattered echo is received. Zoomed view of this received backscattered voltage $V_{echo}$ for capacitive load ($C_L$) of 2 nF and 2.5 nF is shown in Fig. 3(c). The frequency drift is observable for two differing values of capacitive load $C_L$. The anti-resonant frequency of the received echoes for various $C_L$ values are plotted in Figs. 3(d). It can be observed that the anti-resonant frequency monotonically decreases with increasing values of $C_L$. The end-to-end equivalent circuit model simulation is useful not only for the transient analysis but also for evaluating the frequency shift for varying load conditions.

## IV. EXPERIMENTAL SETUP

The experimental setup is shown in Fig. 4 where a Zurich Instruments UHFLI Lock-in amplifier generates the signal to drive the interrogator, and also captures the backscattered echo signal through the T/R switch. The integrated signal analysis tools, arbitrary waveform generator and phase locked loops in this lock-in amplifier enables four different techniques to evaluate the resonance frequency shift of the transducer by varying the capacitive load ($C_L$) by using a custom capacitive termination bank. All measurements are repeated for the same set of capacitive load values.

TABLE I
PARAMETERS USED IN SIMULATIONS

| Parameter | Description | Value | Units |
|---|---|---|---|
| $\rho$ | Transducer density | 7500 | g.m$^{-3}$ |
| $v$ | Transducer wave velocity | 3850 | m.s$^{-1}$ |
| $\varepsilon_{33}$ | Transducer Dielectric constant | 30 | nF.m$^{-1}$ |
| $e_{33}$ | Transducer coupling constant | 23.3 | C.m$^{-2}$ |
| $K_t$ | Transducer coupling factor | 0.5 | - |
| $D$ | Transducer diameter | 14 | mm |
| $T$ | Transducer thickness | 8 | mm |
| $A = \pi (D/2)^2$ | Transducer area | 154 x 10$^{-6}$ | m |
| $C_0 = A\varepsilon_{33}/T$ | Transducer capacitance | 58 | nF |
| $\tau_c = T/v$ | Transducer electrical length | 2 | μs |
| $N = Ae_{33}/T$ | Transducer turn ratio | 0.4483 | C.m$^{-1}$ |
| $h = N/C_0$ | Piezoelectric constant | 7.86 x 10$^6$ | V.m$^{-1}$ |
| $Z_c = \rho vA$ | Transducer radiation impedance | 4445 | Kg.s$^{-1}$ |
| $\rho_a$ | Density of air | 1.2 | Kg.m$^{-3}$ |
| $v_a$ | Wave velocity in air | 343 | m.s$^{-1}$ |
| $D$ | Air gap | 2 | mm |
| $\alpha$ | Air attenuation constant | 0.97 | Np.m$^{-1}$ |
| $R_a = 2\rho_a v_a A\alpha$ | Air lossy line resistance | 124 | mΩ |
| $L_a = A\rho_a$ | Air lossy line inductance | 184 | μH |
| $C_a = 1/A\rho_a v_a{}^2$ | Air lossy line capacitance | 46 | mF |

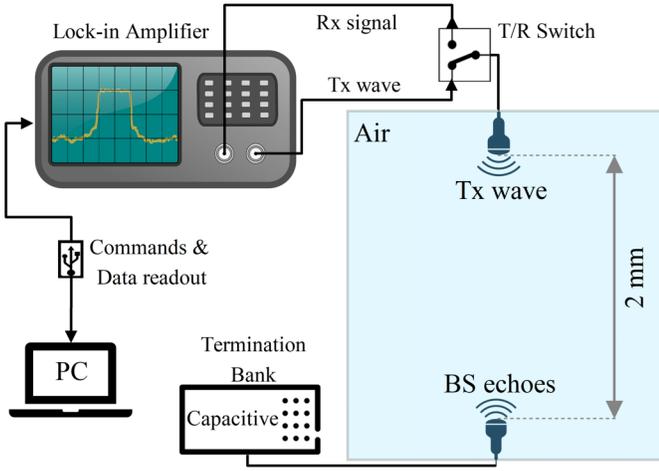

Fig. 4. Experimental setup.

**Ringdown signal analysis:** Similar to the LTSPICE simulations, the interrogator is excited by 10 cycles of sinewave at the parallel resonance frequency. We first focused on ringdown response because it is related to the resonance frequency of the system and damping factor. To capture the ringdown signal, UHFLI's oscilloscope is triggered accordingly and an FFT is performed. The anti-resonance frequency is determined based on this frequency-domain signal.

**Chirp spectroscopy:** To capture the frequency response of the entire system rapidly, the interrogator is excited using a linear chirp signal with a frequency span from 35 to 45 kHz. The chirp is generated by the arbitrary waveform generator and the backscattered signal analyzed using the oscilloscope in FFT mode. This provides the frequency response in less than 1 s. However, this procedure does not provide the phase response of the system.

**Bode analysis:** To obtain both the amplitude and phase of the Tx wave, the lock-in amplifier drives the interrogator with a sine signal and demodulates the resulting Rx wave using the same sine signal as the reference. The frequency response is captured by sweeping the frequency of the driving sine signal and repeating this measurement at each frequency step. The lock-in detection provides the best signal-to-noise ratio among all the measurements, but a single sweep can take more than a few minutes.

**Frequency tracking with a phase locked loop (PLL):** The setpoint is chosen as the phase at anti-resonance frequency, when there is no load attached to the USN. The PLL then adjusts the frequency of the oscillator to match the setpoint value. To track a resonance frequency in a linear system, PLLs are often employed. The PLL integrated in the UHFLI lock-in amplifier, detects the phase using its demodulator. This way, the oscillator frequency follows the anti-resonance frequency of the USN whenever the capacitive load is changed. The frequency is then easily observable for varying values of capacitive load.

## V. RESULTS AND DISCUSSION

Frequency sweeps for varying values of $C_L$ are plotted in Fig.

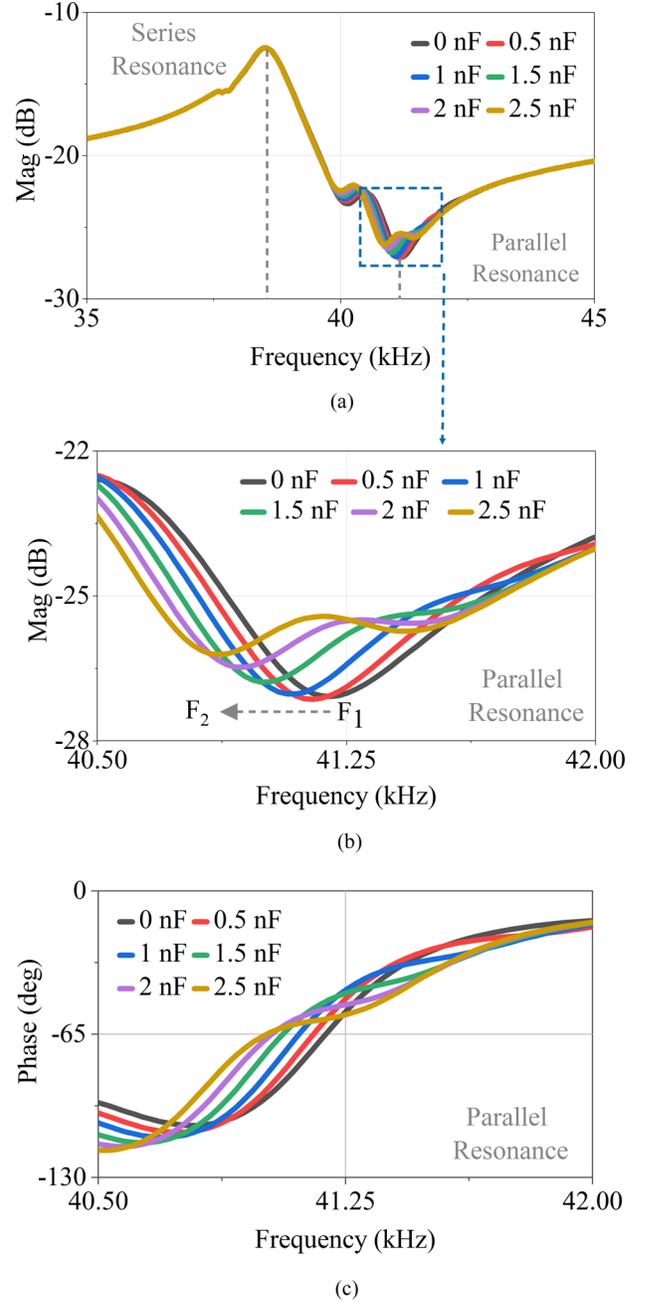

Fig. 5. (a) Frequency response of test setup (b) Zoomed view of magnitude around parallel resonance (c) Zoomed view of phase around parallel resonance

5(a). It can be seen that the series resonance remains unchanged for varying loads ($C_L$). A zoomed view around parallel resonance is shown in Fig. 5(b) and phase values are plotted in Fig. 5(c). It can be seen that the parallel resonance frequency shifts to lower values as the value of $C_L$ is increased. This behavior is consistent with the simulation results of Fig. 3(d), that showed a decrease in parallel resonance frequency vs. increasing $C_L$.

The resonance shift vs. $C_L$ for different measurement types is plotted in Fig. 6. First, the resonance frequency is found when there is no load ($C_L$=0) attached to the USN to be used as reference. Then, $C_L$ is changed, and the shift is calculated by subtracting the new resonance frequency from the reference. It can be seen that the three open-loop measurement types display overlapping behavior. PLL is a closed looped measurement and is slightly diverging from other measurement methods. The reason for this divergence is because PLL works by following the given phase setpoint. In these measurements, we use the setpoint equal to the phase value at parallel resonance frequency when $C_L$=0 i.e., the reference measurement. PLL then follows the new resonance frequency by tracking this setpoint value. Usually, the PLL is used to track resonance, where the phase value does not deviate too much from the reference value. We observed that by changing the capacitive load, the phase value at anti-resonance deviated too much from the reference value. As a result of this, the PLL is not able to track the resonance frequency effectively, due to which the results of PLL differ from the other measurement types.

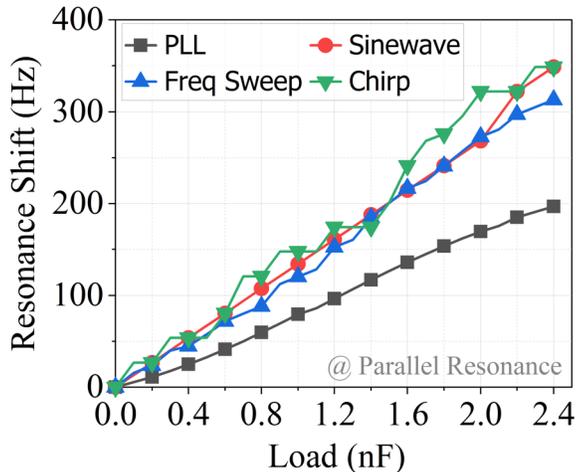

Fig. 6. Parallel resonance frequency shift vs. CL for different measurement types.

The chirp and PLL methods are very fast as compared to sinewave excitation and linear frequency sweep. Chirp is the preferred method when fast measurements are needed with a small compromise on frequency drift values. Overall, linear frequency sweep seems to be most stable method if measurement time can be traded for accuracy.

## VI. CONCLUSION

In this paper, a frequency-based ultrasound backscatter communication method was proposed which could be used for the development of wireless sensor nodes. The simulations and experiments confirm the feasibility of frequency-based sensing in ultrasonic based communication in air domain. Even though the frequency range covered in our studies are limited to airborne applications, the proposed method is also applicable to medical applications with frequencies in MHz range. This method can easily be extended to US implants and frequency-based communication can be used in USNs for airborne and implant applications to overcome challenges posed by amplitude-based sensing.

## REFERENCES

[1] B. M. G. Rosa and G. Z. Yang, "Ultrasound Powered Implants: Design, Performance Considerations and Simulation Results," Sci. Rep., vol. 10, no. 1, pp. 1–16, 2020, doi: 10.1038/s41598-020-63097-2.

[2] D. K. Piech et al., "A wireless millimetre-scale implantable neural stimulator with ultrasonically powered bidirectional communication," Nat. Biomed. Eng., vol. 4, no. February, 2020, doi: 10.1038/s41551-020-0518-9.

[3] F. Mazzilli et al., "In-vitro platform to study ultrasound as source for wireless energy transfer and communication for implanted medical devices," in International Conference of the IEEE EMBS, 2010, pp. 3751–3754, doi: 10.1109/IEMBS.2010.5627541.

[4] H. Kawanabe, T. Katane, H. Saotome, O. Saito, and K. Kobayashi, "Power and information transmission to implanted medical devices using ultrasonic," Jpn. J. Appl. Phys., vol. 40, no. 5 B, pp. 3865–3866, 2001, doi: 10.1143/jjap.40.3865.

[5] M. M. Ghanbari and R. Muller, "Optimizing Volumetric Efficiency and Backscatter Communication in Biosensing Ultrasonic Implants," IEEE Trans. Biomed. Circuits Syst., pp. 1–12, 2020, doi: 10.1109/TBCAS.2020.3033488.

[6] S. Ozeri and D. Shmilovitz, "Simultaneous backward data transmission and power harvesting in an ultrasonic transcutaneous energy transfer link employing acoustically dependent electric impedance modulation," Ultrasonics, vol. 54, no. 7, pp. 1929–1937, 2014, doi: 10.1016/j.ultras.2014.04.019.

[7] L. Beker, "Electro-mechanically optimized ultrasonic implants with backscatter communication," Invention disclosure filed through Koç University TTO, Feb 2021.

[8] W. P. Mason, Electromechanical Transducers and Wave Filters, 2nd ed. New Jersey: D. Van Nostrand Company, 1948.

[9] S. Sherrit, S. P. Leary, B. P. Dolgin, and Y. Bar-cohen, "Comparison of the Mason and KLM Equivalent Circuits for Piezoelectric Resonators in the Thickness Mode," pp. 921–926, 1999.

[10] M. Redwood, "Transient Performance of a Piezoelectric Transducer," J. Acoust. Soc. Am., vol. 33, no. 4, pp. 527–536, 1961.

[11] R. Krimholtz, D. A. Leedom, and G. L. Matthaei, "New Equivalent Circuits for Elementary Piezoelectric Transducers," Electron Lett., vol. 6, no. 13, pp. 398–399, 1970.

[12] W. M. Leach, "Controlled-Source Analogous Circuits and SPICE Models for Piezoelectric Transducers," IEEE Trans. Ultrason. Ferroelectr. Freq. Control, vol. 41, no. 1, pp. 60–66, 1994, doi: 10.1109/58.265821.

[13] D. Zheng, Y. Mao, Z. Cui, and S. Lv, "Optimization of Equivalent Circuit Model for Piezoelectric Ultrasonic Transducer," 2018 IEEE Int. Instrum. Meas. Technol. Conf., pp. 1–5, 2018.

[14] S. Lin, "Effect of electric load impedances on the performance of sandwich piezoelectric transducers," IEEE Trans. Ultrason. Ferroelectr. Freq. Control, vol. 51, no. 10, pp. 1280–1286, 2004, doi: 10.1109/TUFFC.2004.1350956.